\documentclass[
    final            
  ]
  {aipproc}

\layoutstyle{8x11single}

\def\ccpiz{CC$\pi^0$}
\def\che{\u{C}erenkov}

\begin{document}

\title{\ccpiz\ Event Reconstruction at MiniBooNE}


\author{R.H.~Nelson}{
  address={University of Colorado, Dept. of Physics, 390 UCB, Boulder, CO
80309, USA},
  ,email={robert@pizero.colorado.edu}
}

\begin{abstract}
We describe the development of a fitter to reconstruct $\nu_\mu$ induced
Charged-Current single $\pi^0$ (\ccpiz) events in an oil \che\ detector (CH$_2$).
These events are fit using a generic muon and two photon extended track
hypothesis from a common event vertex.  The development of ring finding and
particle identification are described.  Comparisons between data and Monte
Carlo are presented for a few kinematic distributions.
\end{abstract}

\maketitle


\section{Introduction} 
Experimental measurements of the charged-current neutral pion (\ccpiz) mode
are of interest to the community for several reasons.  First, like
charged-current quasi-elastics (CCQE) a \ccpiz\ event only scatters off of
the neutrons in a nucleus.  Second, there is no coherent interaction, so
this channel uniquely probes incoherent pion production.  Third, \ccpiz\
events are a small background for experiments that intend on using CCQE to search for neutrino oscillations.
As these measurements get more precise, a more accurate background prediction
is required.  Finally, this measurement has orders of magnitude more
statistics than previous measurements which will allow for the measurement
of various, previously unmeasured, differential cross-sections. 

This note describes the development of a three-(\che)ring fitter needed for the
measurement of \ccpiz\ differential cross-sections.  The experiment is described along
with the observable signal.  The development of the fitter is detailed, an
event sample is isolated, and various kinematic distributions are shown for
both data and Monte Carlo (MC).

\subsection{MiniBooNE} 
The Mini-Booster-Neutrino-Experiment (MiniBooNE) is described in
detail in Ref.~\cite{boone:detnim}.  The apparatus consists of a pulsed
neutrino beam originating 500 meters away from an 800 ton, spherical, oil \che\ detector.
The neutrino beam is a mostly pure $\nu_\mu$ beam with peak energy of
$\sim$700 MeV with the majority of the flux below 1500 MeV~\cite{boone:flux}.  The detector is
divided into two optically isolated regions; an inner detector region 575 cm
in radius surrounded by 1280 8-inch photo-tubes (11.3\% photo-cathode coverage),
an outer veto region that extends to 610.6 cm in radius monitored by 240
photo-tubes.   

A basic event classification in the MiniBooNE data stream takes advantage of the pulsed
nature of the neutrino beam.  A typical event in the tank produces enough
light during a short time to be detected by some fraction of the photo-tubes.  A
cluster of these hits in a narrow time window is called a ``subevent.''  A
neutrino event may have any number of subevents within the beam time
window.  The prompt neutrino interaction produces one subevent; subsequent
stopped muon decays produce additional subevents.  For example, muon-CCQE
events differ from electron-CCQE events in that they ideally have 2
subevents (one from the initial neutrino interaction, the second from the
electron from the muon decay), where an electron-CCQE should have only one
subevent.  These idealizations can break down ($\mu^-$ capture, a coincident cosmic
ray, a muon decay during the first subevent, etc.), nevertheless as a basic
classifier, dividing events by the number of subevents separates out the
various classes of events.   A \ccpiz\ event is expected to have 2
subevents.  The largest background, CC$\pi^+$ events, are expected to have 3
subevents. 

The MC simulations of pion production in the MiniBooNE detector are based on
the Rein-Sehgal model~\cite{model:RS} with an $M_A=1.1\;$GeV~\cite{nuance}. The Smith-Moniz
formalism~\cite{model:SM} is used for CCQE production with parameters tuned to reproduce the
MiniBooNE CCQE data~\cite{boone:ccqe}.

\subsection{Observable \ccpiz\ events} 
To avoid the added complication of trying to extract cross-sections at the
nucleon level, an observable signal is defined.  The goal is to avoid the
model dependencies associated with trying to extend what is experimentally
observable (particles that emerge from the target nucleus) to what occurred
at the initial neutrino interaction. The ``observable \ccpiz'' signal is
defined as a muon plus a $\pi^0$ that exited the target nucleus (carbon)
with no other mesons.  This not only includes ``true'' \ccpiz\ events
(70\%)~\cite{nuance}, but also CC$\pi^+$ events that charge-exchanged
($\pi^+\rightarrow\pi^0$) within the target nucleus (21\%), CCQE events
(6\%), and a remainder of NC and CC multi-pion events.   It is important to
note that any $\pi^0$ created outside of the initial target nucleus, through
$\pi^+$ and nucleon reinteractions in the mineral oil, is not considered an
observable \ccpiz\ event.  These events are referred to as ``tank-$\pi^0$'' events. 

\section{Event reconstruction} 
MiniBooNE incorporates an extended-track, maximum likelihood method for
event reconstruction~\cite{boone:recnim}.  A track is characterized by 7 parameters:
the event vertex
($x,y,z$), the time ($t$), direction ($\theta,\phi$), and kinetic energy ($E$).
The \che\ and scintillation light produced by a track are separated into
prompt and late portions.  The generated light is then propagated through
the mineral oil via an extensive optical model until the light reaches the 
photo-tubes.  For each PMT, a probability is assigned to address whether
the tube was hit, the time of the hit, and the resulting deposited charge.
These probabilities are compared with the measured charge and time
associated with each tube.  The hit times are dominated by the \che\ light.
The charges are simply the sum of the scintillation and \che\ portions.  The
track parameters are varied and the maximum likelihood is chosen to best
describe the track parameters.  In practice, the minimum of the negative
logarithm of the likelihood is used.  This way the time and charge
likelihoods can be calculated separately and then added together to get the
total likelihood.  The light profiles for different types of particles (in
this case $\mu$ or $e/\gamma$\footnote{The difference between electrons and
photons is that photons travel some distance before they produce a track
similar to electrons.}) are used to distinguish between event types.

For multiple tracks, the likelihoods for each track can be combined.  The
charge likelihoods can be simply added together.  The time likelihoods are
weighted conditionally by the closest track (defined by the track's
mid-point), then added together.  The number of parameters for multi-track
events depend on the event type.  For a \ccpiz\ event hypothesis, it is
assumed that the $\pi^0$ decays to 2 photons immediately\footnote{Branching
fraction of $\pi^0\rightarrow\gamma\gamma$ is 98.8\%~\cite{pdg}.}.  The signature of
these events is a muon and 2 photons that come from a common vertex.
Therefore, there are 15 parameters to fit: the event vertex ($x,y,z$); the
time ($t$); the muon's energy and direction ($E_\mu,\theta_\mu,\phi_\mu$); and
the photons' energy, direction, and conversion length
($E_i,\theta_i,\phi_i,s_i$ where $i=\gamma_1,\gamma_2$).  

Tracks are searched for in a step-wise fashion.  A one-track fit is
performed, which finds one of the three tracks.
While keeping that track fixed in the likelihood function, a scan of 400
points in solid angle with a 200 MeV (chosen at the peak of the photon
energy spectrum) second track about the event vertex is performed.  The best
likelihood from the scan is chosen, and all 
parameters (except the conversion lengths) are allowed to float for the two
tracks.  The two tracks are then fixed in the likelihood function and the
full solid angle is scanned for a third track.  The best likelihood is then used
as a seed for a three-track fit (again, no conversion lengths).  Once the
event vertex, track directions, and rough estimates for the track energies
are found, three fits are then performed in parallel.  These fits swap out
two of the tracks for photons seeded with conversion lengths of 50
cm (roughly the radiation length in oil) This is done for
the three possible particle type configurations.  One of the three fits then
gets selected as the final fit based on the fit likelihood and whether the
muon track points toward the electron vertex in the second subevent.

\subsection{Cuts} 
Due to the long processing time per event, the two subevent sample must be
reduced to
a more manageable size before this fitter is run.  Since various one-track
fits were already run on the sample, they can be used reduce the CCQE
contamination.  Most of the events in the sample are muon-CCQE.  This implies that
they should look more like a muon than an electron.  Electrons tend to
produce fuzzy rings while muons are more rigid.  To a fitter that only fits
single electrons and muons, a multiple track event would look more like an
electron.  A cut is applied that selects more electron-like events in the fit
likelihood.  This cut defines the sample that will be fit by the \ccpiz\
fitter.  It is 88\% efficient at keeping observable \ccpiz\ events.  This
reduces the two-subevent sample from $\sim$250,000 to only $\sim$43,500 events
that need to be fit.  After the events are fit they are again selected to isolate
a purer observable \ccpiz\ sample.  Two cuts are applied; a fit likelihood
cut that removes events without $\pi^0$, and a cut on the smallest angle
between the three reconstructed tracks removing events with poorer particle
ID.  Eventually, a $\pi^0$ mass cut will be applied to reduce some
backgrounds further, and to select well reconstructed observable \ccpiz\
event.  However, no mass cut has been applied in this analysis.

Table~\ref{cuts} shows the effects of these cuts.  The two subevent sample
is mostly CCQE with only a 6.5\% observable \ccpiz\ purity.  After all the
cuts, the sample is reduced by 75\%, however, the \ccpiz\ purity increases to
46.4\%.  Including the two subevent cut (removes $\sim$60\% of observable
\ccpiz\ events), this factors in as about a 12\% 
total efficiency for \ccpiz\ events.  The CCQE events are reduced to less
than 1\% efficiency but still comprise 10\% of the final sample.
Unfortunately, the tank $\pi^0$ are kept in roughly the same proportions as
the signal.  They mostly come from CC$\pi^+$ events that charge exchange in
the mineral oil and are the largest background.  The remaining tank $\pi^0$
background are mostly from CC multi$\;\pi$, NC events, and DIS.

\begin{table}
\begin{tabular}{l|rc}
\tablehead{1}{c}{b}{MC Sample}                  &
\tablehead{1}{r}{b}{2SE + tank + veto}          &
\tablehead{1}{r}{b}{+ event filter + likelihood + small angle} \\
\hline
CCQE              & 155480 (\phantom{0}63\%) &  \phantom{0}0.6\% (\phantom{0}10\%) \\ 
\ccpiz            &  15615 (\phantom{00}6\%) &  19.3\% (\phantom{0}34\%) \\ 
CC$\pi^+$         &  62167 (\phantom{0}25\%) &  \phantom{0}1.4\% (\phantom{0}36\%) \\ 
everything else   &  14764 (\phantom{00}6\%) &  28.3\% (\phantom{0}20\%) \\
\hline
Observable \ccpiz &  16098 (\phantom{00}6\%) &  25.7\% (\phantom{0}47\%) \\ 
tank $\pi^0$      &  15507 (\phantom{00}6\%) &  25.5\% (\phantom{0}44\%) \\ 
everything else   & 216421 (\phantom{0}87\%) &  \phantom{0}0.4\% (\phantom{00}9\%) \\
\hline
total             & 248026 (100\%)           &  \phantom{0}3.6\% (100\%)                  
\end{tabular}
\caption{This table is divided into two sections.  The
first section divides the sample into the absolute modes from the {\tt
NUANCE}~\cite{nuance} predictions.  The second
section divides the sample into the observable \ccpiz\ signal, and the tank
$\pi^0$.  The first column shows the predicted event numbers with the basic
cuts along with the purity in parentheses.  The second column adds the event
filter and analysis cuts and shows the efficiencies relative to the first
column along with the purity in parentheses.  The final sample is divided
between signal and background with a 26\% signal efficiency.  
}
\label{cuts}
\end{table}

\section{Kinematics} 
The usefulness of this fitter is ultimately measured by how well it can be
used to reconstruct various physics parameters of the event.  To that end, it
is useful to understand how reconstructing the muon and the two photons can
be used to infer the properties of the incident neutrino and the intermediate
states.  A \ccpiz\ event is of the form $\nu_\mu
n\rightarrow\mu^-p\pi^0\rightarrow\mu^-p\gamma\gamma$. Conservation of 4-momentum yields
\begin{equation}
(p_\nu + p_n - p_X)^2 = m_p^2
\end{equation}
where $p_X\equiv p_\mu + p_{\gamma1} + p_{\gamma2}$, replaces the typical lepton
momentum used to derive the standard CCQE neutrino energy formula.  One
particle is replaceable by its invariant mass, in this case the unmeasured
proton.  Since the fitter measures $p_X$, the neutron is assumed to be
at rest, the neutrino is assumed to
travel in the beam direction, then the neutrino energy is fully specified.
The neutrino energy is given by
\begin{equation}
E_\nu = \frac{m^2_p-m^2_n-m^2_X +2m_nE_X}{2(m_n - E_X+|\mathrm{\mathbf p}_X|\cos\theta_{\nu X})}
\end{equation}
where properties of $X$ are measured in the lab frame.
Now that all the initial and final state particles have been fully
specified, the  $\pi^0$, $Q^2$, and the $\Delta$ resonance internal states
can be reconstructed.  However, no assumption is made that this
is actually a $\Delta$ resonance (i.e. no mass assumption), it can be any of the 17
resonances in the MC, though the $\Delta$ is by far the most prominent
source at MiniBooNE energies.  They are reconstructed through
\begin{eqnarray}
p_\pi    & = & p_{\gamma1} + p_{\gamma2} \\
Q^2      & = & -(p_\nu - p_\mu)^2 \\
p_\Delta & = & p_\nu - p_\mu + p_n. 
\end{eqnarray}

\begin{figure}
\label{mu:fig}
  \includegraphics[scale=.4]{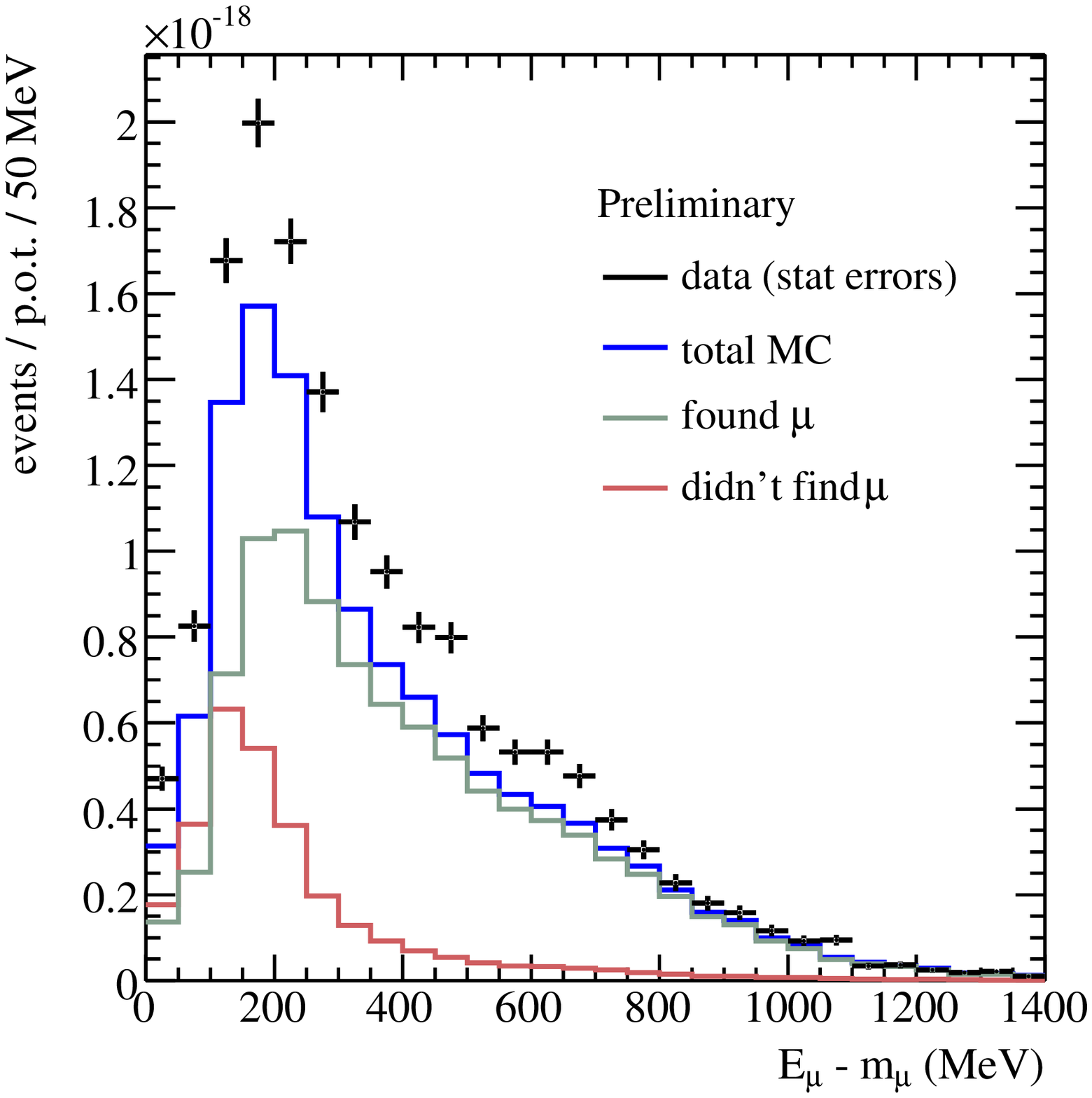}
  \includegraphics[scale=.4]{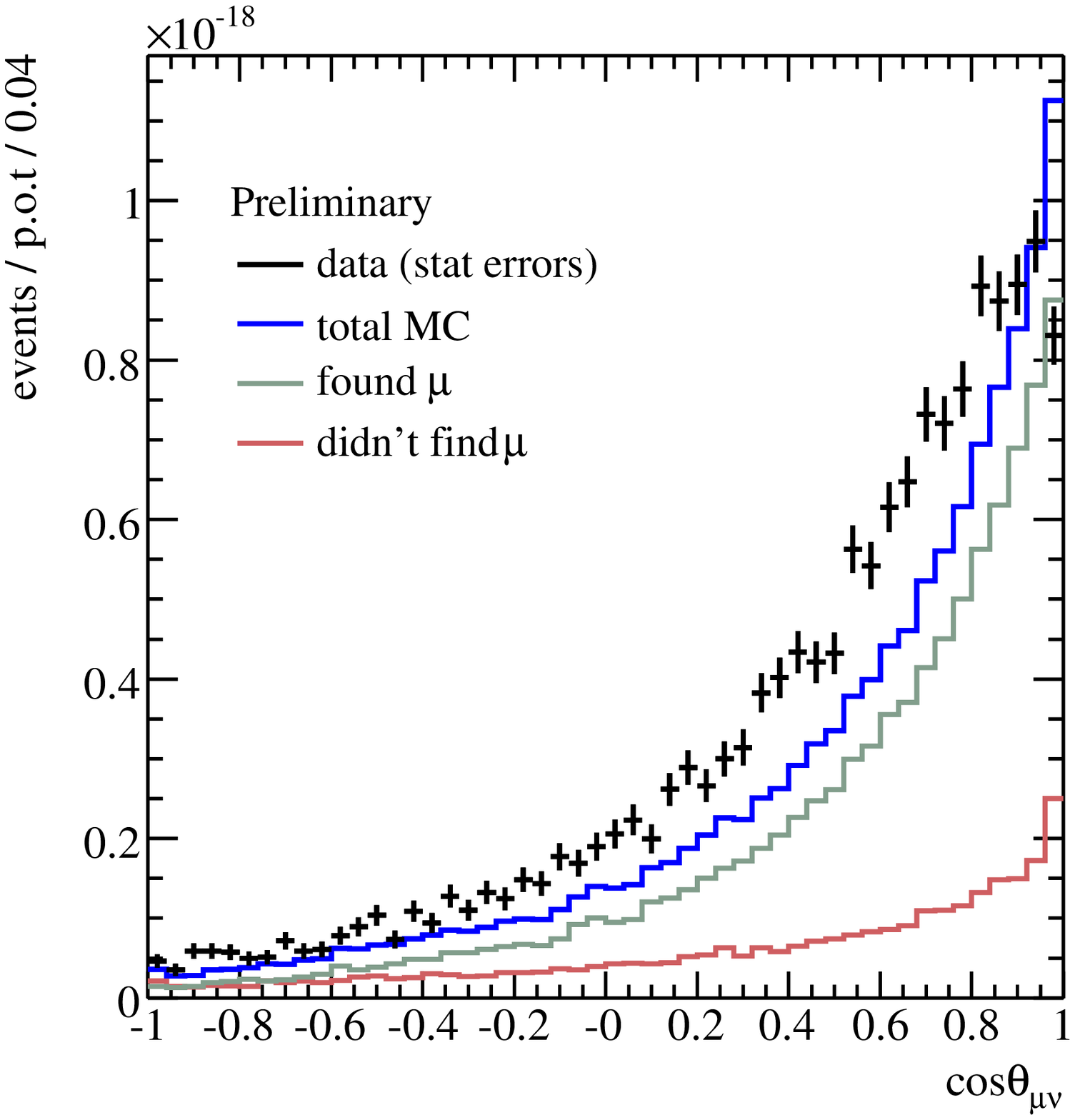}
  \caption{The reconstructed muon kinetic energy (left) and the muon angle
relative to the neutrino beam (right).  The MC is broken up into two samples, one where
the particle ID correctly identified the muon (grey), and one where it did
not (red).  Data is normalized by proton on target (p.o.t.).  MC is
normalized by the default {\tt NUANCE} cross-section predictions. }
\end{figure}

\begin{figure}
\label{pi:fig}
  \includegraphics[scale=.4]{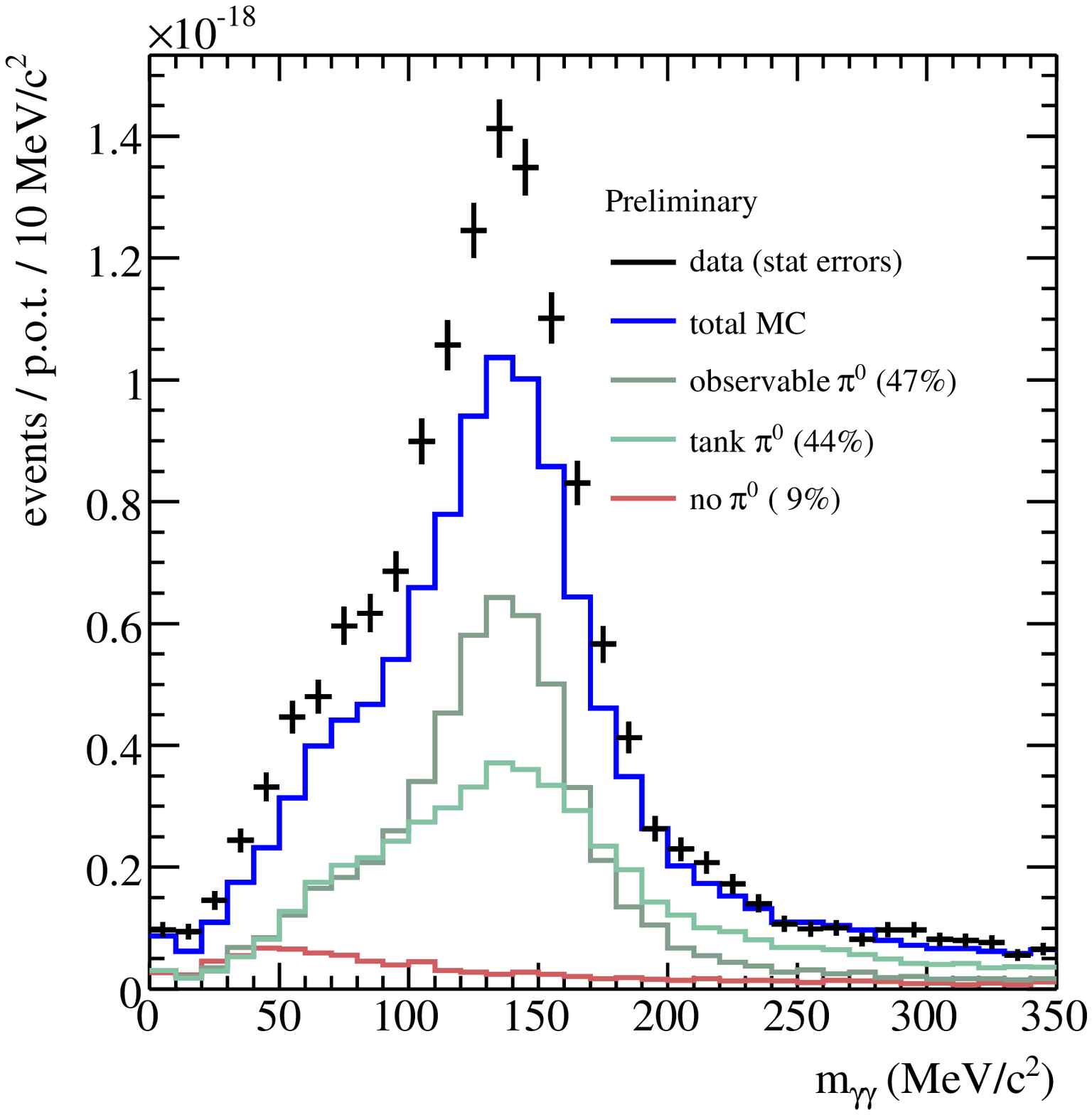}
  \includegraphics[scale=.4]{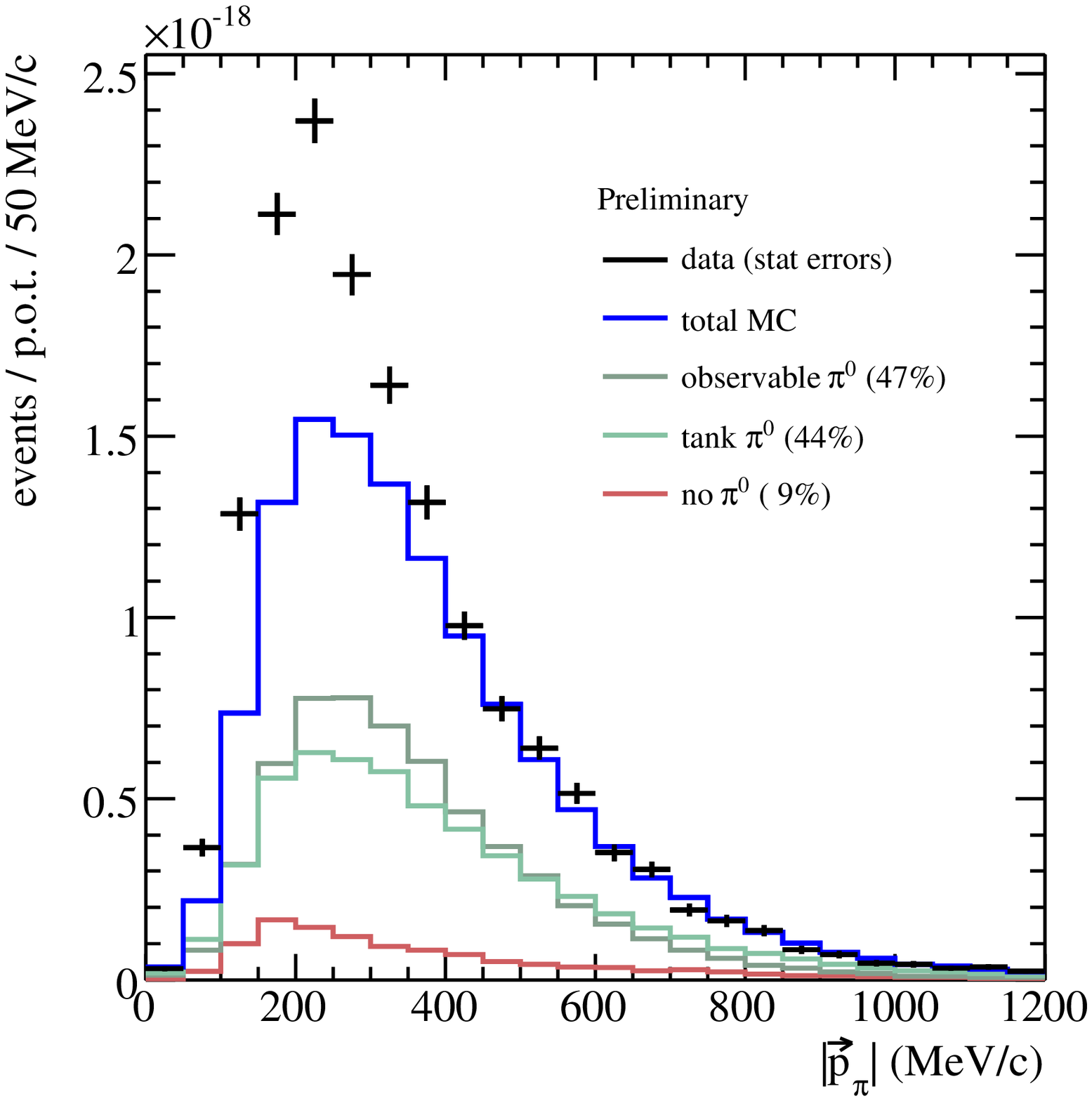}
  \caption{The reconstructed $\pi^0$ mass (left) and 3-momentum (right).  The
MC is divided into three samples; observable \ccpiz\ (grey), tank $\pi^0$
(green), and events without a $\pi^0$ (red).  The mass plot shows a nice peaks at
the known $\pi^0$ mass for both the observable and tank $\pi^0$ cases.  Each
sample is normalized by proton on target (p.o.t.).}
\end{figure}

\begin{figure}
\label{eneu}
  \includegraphics[scale=.4]{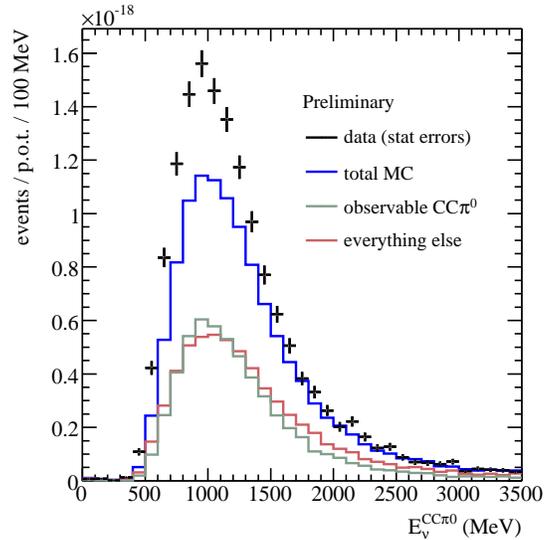}
  \caption{The reconstructed neutrino energy.  The MC sample has been
divided into observable \ccpiz\ (grey) and everything else (red).  Each
sample is normalized by proton on target (p.o.t.).} 
\end{figure}

\begin{figure}
\label{q2}
  \includegraphics[scale=.4]{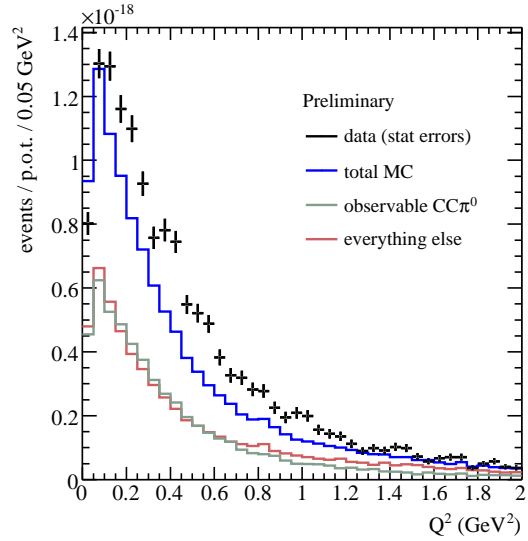}
  \caption{The reconstructed 4-momentum transfer, $Q^2$.  Each
sample is normalized by proton on target (p.o.t.).}
\end{figure}

While the MC appears to underpredict the number of \ccpiz\ events observed
in data\footnote{The normalization difference is consistent with the
differences observed in other MiniBooNE CC samples.},
most distributions agree fairly well in shape. Fig.~\ref{mu:fig} shows the
reconstructed muon kinetic energy and muon angle relative to the beam.  Data and MC
agree quite well (in shape) with a muon energy resolution of 9\%.
Fig.~\ref{pi:fig} shows the reconstructed pion mass and 3-momentum. The mass
plot is striking because of how well the three-track fit matches the known
$\pi^0$ mass when no $\pi^0$ mass assumptions were used during the fit.
Both the mass and the momentum match well in shape. 
Fig.~\ref{eneu} shows the reconstructed neutrino energy.  The neutrino
energy matches well even for modes without $\pi^0$.  This is partly because
the neutrino energy is roughly correlated with the total energy deposited in
the tank and the fitter does a good job on measuring the total visible
light.  The neutrino energy resolution is found to be 11\%.  Finally,
Fig.~\ref{q2} shows the measured 4-momentum transfer, $Q^2$.  Differences in
these distributions probe the model used to predict single pion modes.  

\section{Conclusions} 
A method for reconstructing \ccpiz\ events in a \che\ detector has been
described.  After a reduction of the data to increase the signal purity, a
three-track fit was performed. The sample was then further reduce to isolate
a purer sample of observable \ccpiz\ events.  Since this fitter reconstructs
both the muon and the $\pi^0$ in an event, it has the ability to measure
many differential cross-sections.  Among the possible 
measurements will be: $\sigma(E_\nu)$, $\frac{d\sigma}{dE_\mu}$,
$\frac{d\sigma}{\cos\theta_\mu}$, $\frac{d\sigma}{d|\vec{p}_\pi|}$,
$\frac{d\sigma}{\cos\theta_\pi}$, $\frac{d\sigma}{dQ^2}$.  Hopefully, this
suite of measurements will aid in the understanding of CC single-pion
production on nuclear targets. 



\begin{theacknowledgments}
The author would like to acknowledge the MiniBooNE experiment without which
this work would not have been possible.  Also, the University of Colorado in
Boulder, Fermilab, The Department of Energy, and the National Science
foundation.  Finally, to the organizers of the Sixth International Workshop on 
Neutrino-Nucleus Interactions in the Few-GeV Region (NuInt 2009).

\end{theacknowledgments}

\bibliographystyle{aipproc}   

\bibliography{nuintproc}

\end{document}